\documentclass[final]{elsart}

\usepackage{natbib,graphicx,amssymb}

\newcommand{\ie}{{\it i.e.}}

\begin{document}
\begin{frontmatter}



\title{Variations of the Solar Acoustic High-Degree Mode Frequencies over
Solar Cycle 23} 


\author[su]{M.C.~Rabello-Soares},
\author[cfa]{Sylvain~G.~Korzennik},
\author[su]{J.~Schou}
\address[su]{W. W. Hansen Experimental Physics Laboratory, 
	Stanford University, 455 Via Palou, Stanford, CA 94305, USA}
\address[cfa]{Harvard-Smithsonian Center for Astrophysics,
              60 Garden St, Cambridge, MA 02138, USA}

\begin{abstract}

Using full-disk observations obtained with the Michelson Doppler Imager (MDI)
on board the Solar and Heliospheric Observatory (SOHO) spacecraft, we present
variations of the solar acoustic mode frequencies caused by the solar activity
cycle.  High-degree (100 $< \ell <$ 900) solar acoustic modes were analyzed
using global helioseismology analysis techniques over most of solar cycle 23.
We followed the methodology described in details in \citet{KRS04} to infer
unbiased estimates of high-degree mode parameters \citep[see also][]{RKS06}. We have
removed most of the known instrumental and observational effects that affect
specifically high-degree modes.
We show that the high-degree changes are in good agreement with the
medium-degree results, except for years when the instrument was highly
defocused. We analyzed and discuss the effect of defocusing on high degree
estimation.
Our results for high-degree modes confirm that the frequency shift scaled by
the relative mode inertia is a function of frequency 
and it is independent of degree.
\end{abstract}


\end{frontmatter}

\section{Introduction}

The correlation between solar acoustic mode frequencies and the magnetic
activity cycle is well established and has been substantially studied during
the last and current solar cycles \citep[see for example][and references
therein]{DG05}.  The general frequency shift behavior suggests that the
perturbation by the solar cycle on the acoustic frequencies occurs very close
to the solar surface.  The physical mechanism, however, is still a matter of
debate, namely whether frequencies are changed directly by
the magnetic field or indirectly through an associated change in the solar
structure (like a pressure change).  A detailed analysis of the frequency
shift characteristics will hopefully help understand its physical origin.

We present here a new analysis of high degree mode frequency shifts over most
of solar cycle 23, using a global helioseismology technique.  High-degree
modes are best suited for studying the frequency shift correlation with the
solar cycle since the causing mechanism is believed to be localized near the
surface and high-degree modes are also confined to a region near the
surface. As a result the frequency shifts for high degree modes are up to an
order of magnitude larger than for medium degree modes.

\section{Global Helioseismology Analysis of High-Degree Modes}

In the traditional global helioseismology data analysis methodology, a time
series of full-disk Doppler solar images is decomposed into spherical harmonic
coefficients, characterized by its degree ($\ell$) and its azimuthal order
($m$).  Each coefficient time series is Fourier transformed, and the order of
the radial wavefunction ($n$) gets separated in the frequency domain.
However, a spherical harmonic decomposition is not orthonormal over less than
the full sphere --\ie, the solar surface that can be observed from a single
view point-- resulting in what is referred as spatial leakage.  At low and
intermediate degrees, these leaks are separated in the frequency domain from
the target mode and individual modes can be identified and fitted. However, at
high degrees, the spatial leaks lie closer in frequency (due to a smaller mode
separation) and, at high frequency, they become wider (as the mode lifetimes
get smaller), resulting in
the overlap of the target mode with the spatial leaks that merges individual
peaks into ridges.
The characteristics of the resulting ridge (central frequency, amplitude,
etc\ldots) do not correspond to those of the underlying target mode. This has
so far hindered the estimation of unbiased mode parameters at high degrees.

Our methodology to recover the mode characteristics consists in generating and
fitting a sophisticated model of the underlying modes that contribute to the
ridge power distribution and deduce the offset between the ridge properties
and the underlying target mode \citep{Korzennik98}.  To do this we need a very
good model of the relative amplitude of all the modes that contribute to the
ridge, \ie\ the leakage matrix, which in turn requires a very good knowledge
of the instrumental properties.  The instrumental characteristics must
therefore be very well understood and very precisely measured
\citep[see][]{RKS01}.  Although, the highly successful MDI instrument is
responsible for great progresses in our understanding of the Sun, there remain
some instrumental effects that there are not well characterized and/or
not taken into account in the data analysis \citep{KRS04}.

  We have incorporated most of the known instrumental effects important to the
high-degree analysis in our most recent spatial decomposition, introducing one
effect at a time to estimate the resulting correction on mode parameters
\citep{RKS06}.  The following known instrumental effects were taken into
account in this latest spherical harmonic decomposition, whose results are
presented here: (1) the correct instantaneous image scale, (2) the radial
image distortion, (3) the effect of a tilt of the CCD with respect to the
optical axis, (4) the effective $P$ angle and (5) a correction to the
Carrington elements.

Since we are focusing our attention in this paper on changes of mode
parameters with epoch, we are primarily concerned with instrumental effects
that change over time.
  Although the MDI instrument has been very stable over the years, continuous
exposure to solar radiation has increased the instrument front window
absorption resulting in a continuous small increase of the instrument best
focus. Moreover the change of the front window temperature due to the
satellite orbit around the Sun also adds a small annual variation.  The
instrument has however an adjustable focus with 9 possible
positions\footnote{One focus step corresponds to approximately a third of a
wave} chosen to best suit the science team needs, resulting unfortunately in
abrupt jumps every time that it is changed. These changes are responsible for
the largest variations.  The best focus is periodically and empirically
determined from intensity images taken with different focus positions
\citep{Kuhn04}.
The change in the image scale resulting from a change in focus was shown to be
an important effect \citep{KRS04} and it is taken care in the analysis
presented here.  However, the effect of the amount of defocus was not taken
into account.  A preliminary analysis indicated that it has a small effect on
the frequencies \citep{KRS04}; we have further studied its influence on the
ridge frequency.

\section{Description of the Data Set Used}

Full-disk Dopplergrams, acquired by the MDI instrument while operating in its
$4''$ resolution mode, were used for this analysis. The MDI instrument is
operated in this mode, known as the {\em Dynamics Program}, every year for two
to three months, when the required telemetry bandwidth is available.  The
spherical harmonic decomposition of these Dopplergrams was computed for every
tenth $\ell$ between $100$ and $900$, and the resulting time series Fourier
transformed in shorter segments 
\citep[as described in][namely 4096 minutes or 2.8 days long]{KRS04}
whose power spectra were averaged to produce
an averaged power spectrum\footnote{
Henceforth the number of averaged spectra varies with epoch, from 12 segments in 2003, to 32 in 1997.} 
with a low but adequate frequency resolution to fit
the ridge while reducing the realization noise. Table~\ref{table1} lists the
properties of each epoch included in this study.
The peaks in 
each ($\ell,m$) spectra were then fitted with 
asymmetric Lorentzian profiles and the $2\ell+1$
frequencies for each ($n$, $\ell$) multiplet expanded in
terms of Clebsch-Gordan coefficients up to $a_6$
\citep[see][for details]{KRS04}.

We also compare our results (hereafter referred to as the high-$\ell$ set) to
results obtained by one of us using 
%
the 
MDI
{\em Structure Program}. In
this mode the original Dopplergrams are convolved, on board the MDI
instrument, with a Gaussian and subsampled on a $200 \times 200$ grid, thus
reducing the telemetry requirements but also limiting the spatial resolution
to modes with degree $\ell~<~300$.  The data sets consist of 72-day-long time
series that partially overlap with the {\em Dynamics} periods (see
Table~\ref{table1}).  The central frequencies and the splitting coefficients
were determined directly from fit to symmetric Lorentzians \citep{Schou99}.
Although the instrumental effects listed in Section~2 were not taken into
account here, their effects are much smaller because the spatial leaks are
well separated from the target mode \citep[]{Schou01}.  

\begin{figure}[p]
\begin{center}
\includegraphics[width=10cm]{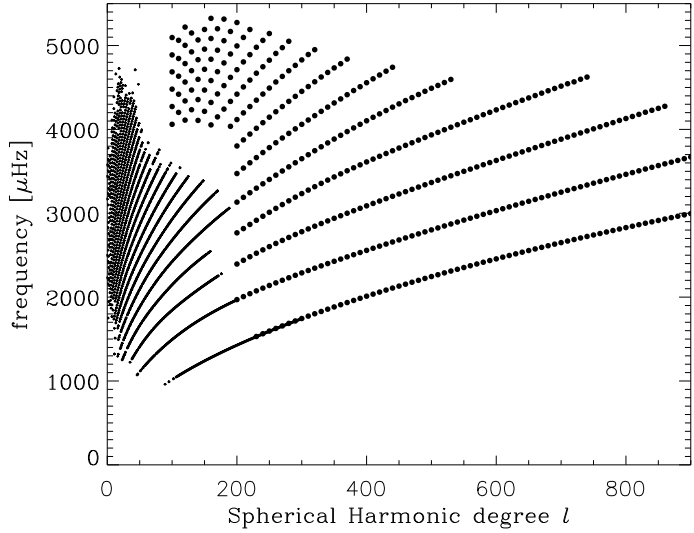}
\end{center}
   \caption{ Coverage, in an $\ell-\nu$ diagram, of the medium-$\ell$
(crosses) and high-$\ell$ (circles) set
for 1996. The coverage is very similar for all epochs.
Modes with $\ell < 200$ and $\nu < 4$ mHz may not completely overlap
with their spatial leaks and were excluded from the present high-$\ell$ analysis.
}
\label{lnu}
\end{figure}

The coverage of both analyses is illustrate in an $\ell-\nu$ diagram in
Fig~\ref{lnu}.  Note that there is a small overlap in the coverage of both
methods. Indeed, by degrading the frequency resolution one can merge resolved
modes into a smooth ridge and apply the ridge fitting methodology developed
for high degrees to intermediate degrees.

\begin{figure}[p]
\begin{center}
\includegraphics[width=10cm]{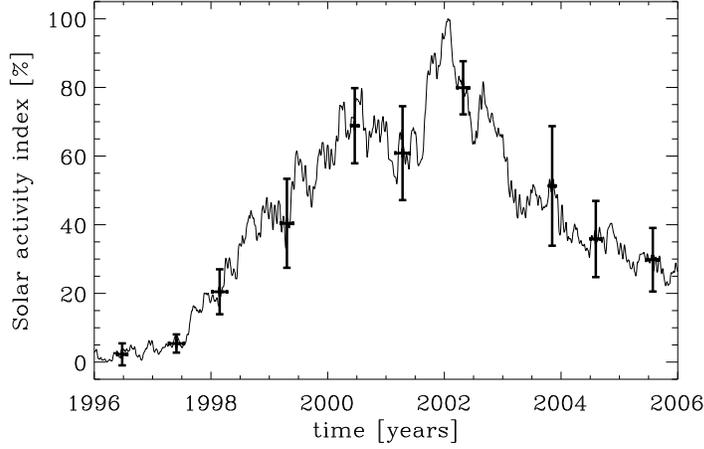}
\end{center}
   \caption{
Solar UV spectral irradiance (given by the NOAA Mg II core-to-wing ratio)
30-day running mean relative to its maximum during solar cycle 23
(line) and its average for each epoch.
}
\label{solaract}
\end{figure}

The MDI data analyzed here, see Table~\ref{table1}, covers much of solar cycle
23, as illustrated in Fig.~\ref{solaract}.  The average solar UV spectral
irradiance (given by the NOAA Mg II core-to-wing ratio) relative to the 30-day
running mean maximum of solar cycle 23 (reached in January 2002) for each
epoch is listed in Table~\ref{table1}. That table also lists the average
defocus of the MDI images during each {\em Dynamics} epoch, in units of MDI
focus steps.  Early on (1996 to 1998) the instrument was set to a large
defocus (up to 2.5 focus steps), while later on the instrument was set to be
nearly on focus (\ie, a defocus of 0.5 focus steps in average).

\section{Analysis of the Frequency Shift} 

\begin{figure}[p]
\begin{center}
\includegraphics[width=14cm]{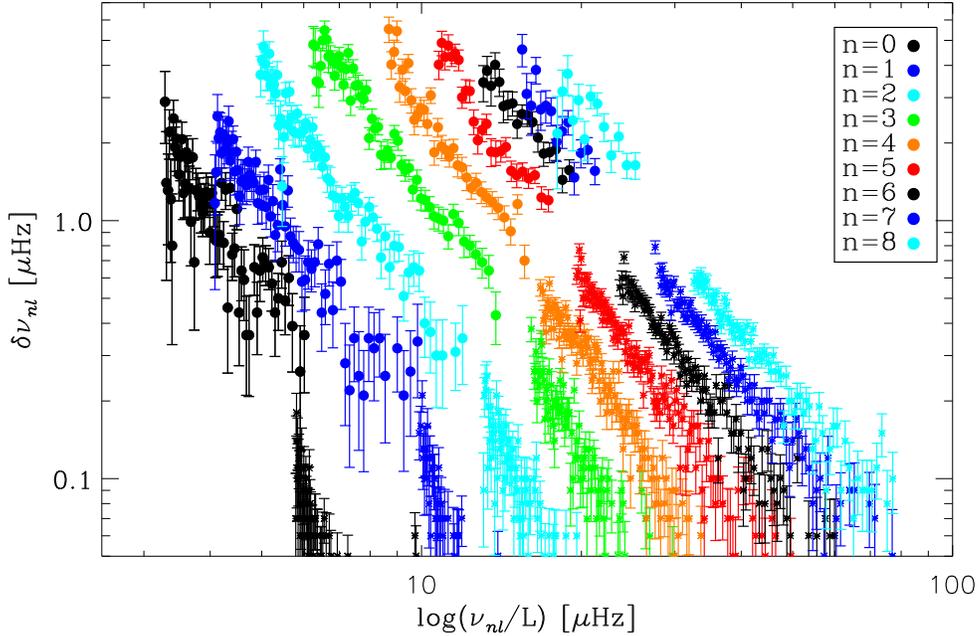}
\end{center}
   \caption{ Differences between the mode frequencies as observed in 2002,
near solar maximum, and in 2005, during moderate
activity.  Only modes with $n=0$ to $n=8$ are plotted
(evens in black, odds in grey).
The medium-$\ell$ set is indicated by stars (lower part
of the plot) and high-$\ell$ set by circles (upper part).  }
\label{freq-ml}
\end{figure}

   Figure~\ref{freq-ml} shows mode frequency differences between 2002 and
2005, namely between high and moderate
activity epochs respectively, resulting from
medium and high degree determinations. 
The frequency shifts for each order, $n$, scale as a power law with the
quantity $\nu/L$ (where $L = \ell + 0.5$, a good proxy of the mode inner
turning point), and can be described in a first approximation by
$\delta\nu_{n,\ell} \propto I^{-1}_{n,\ell}$, where $I_{n,\ell}$ is the mode
inertia \citep[as first noticed by][]{LW90}.

For the observational periods where the MDI instrument was set to a large
defocus (1996 to 1998), our measured frequency shifts for high-$\ell$ do not
agree with results from medium-$\ell$.  The defocus has a substantial effect
on the ridge central frequencies,
contrary to our initial expectations.  The resulting instrumental point spread
function (PSF) is affected and the effect of a varying PSF has not yet been
incorporated in our analysis, therefore we will not include these epochs in
our analysis of changes of mode parameters with the solar cycle.

\begin{figure}[p]
\begin{center}
\includegraphics[width=10cm]{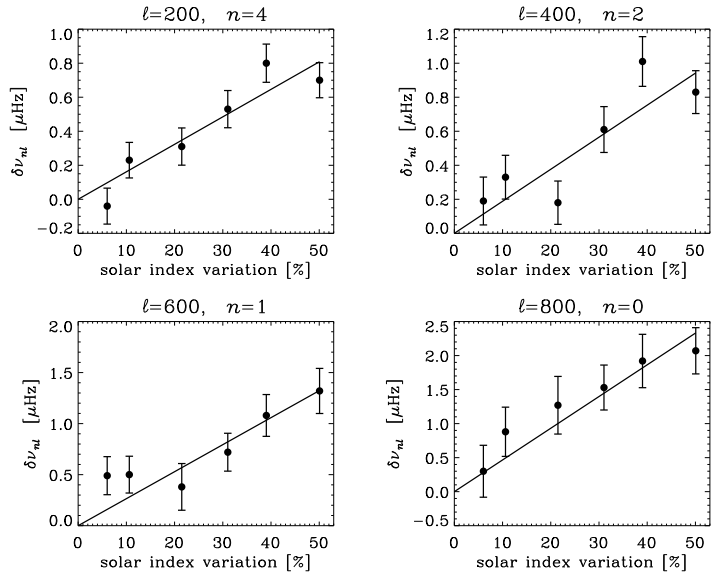}
\end{center}
   \caption{ 
Examples of correlation
between the frequency shift and the solar cycle for six epochs
(1999 to 2004) with respect to 2005
for four high-$\ell$ modes with different degree and order but similar
frequency ($\sim$ 3 mHz).  The error bars are the standard deviation of the
fitted frequencies.  The straight line is a weighted least-squares fit to the
data, keeping the intercept to zero.  }
\label{corr-freq}
\end{figure}

Figure~\ref{corr-freq} shows the correlation between the frequency shift and
the solar activity index variation (given by the solar UV irradiance) for four modes
with frequency around 3 mHz at high degree ($\ell=200, 400, 600$ \& $800$) for
the 1999 to 2004 epochs with respect to the 2005 epoch.
We have fitted, using a weighted least-squares minimization, the frequency
shift to the solar index variation, assuming a linear relationship with a
zero intercept (solid line).  This fit was performed for each ($n$, $\ell$)
mode of the medium and high-$\ell$ sets separately,
resulting in a total of 1468 and 423 
regression fits
respectively.

We also computed correlations between the frequency shift and the Magnetic
Plage Strength Index (Mt.\ Wilson Observatory), the solar radio 10.7 cm flux
(National Research Council of Canada) and the sunspot number (SIDC, RWC
Belgium) using the high-$\ell$ set.  The magnetic index gives similar Pearson
correlation coefficients as the UV flux while the radio flux coefficients are
slightly smaller (by a few percent).
The sunspot number correlation coefficients are also similar to those for the
UV flux but only for frequencies around 3 mHz, 
their correlation coefficients increase as the frequency decreases, by some 20\%.

The frequency shift between solar maximum and minimum can be estimated by
multiplying the slope of the linear fit (solid line in
Fig.~\ref{corr-freq}) by the corresponding solar index variation.  We used
the 30-day running mean values to estimate the minimum and maximum, reached in
March 1996 and January 2002 respectively.  We rejected modes whose Pearson
correlation coefficient is smaller than 0.8, or whose slope uncertainty is
larger than $20\%$ and only fitted modes for which we have estimates for at
least five of the six epochs.
This estimated minimum-to-maximum frequency shift, $\delta\nu^e$, incorporates
information from all the data sets (1999 to 2005) in a consistent way,
decreasing ipso facto the uncertainties.

\begin{figure}[p]
\begin{center}
\includegraphics[width=10cm]{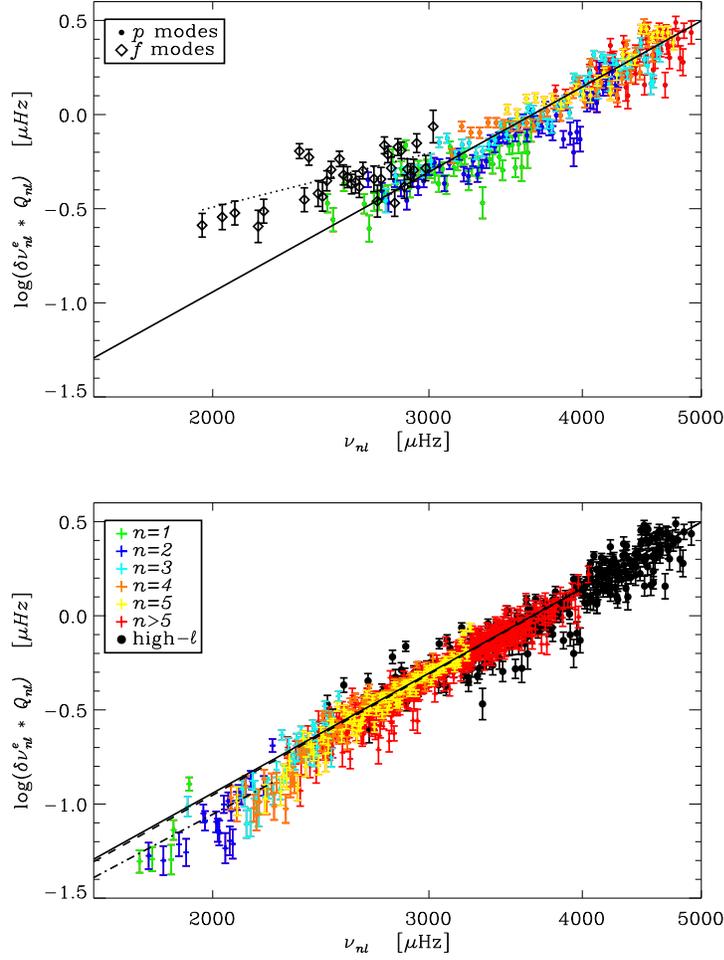}
\end{center}
   \caption{ Minimum-to-maximum frequency shift multiplied by the normalized
mode inertia as a function of frequency.  The upper panel shows only the
high-$\ell$ set and, in the lower panel, the high-$\ell$ $p$ modes are
indicated in black (circles) and the medium-$\ell$ $p$ modes in gray (crosses).
}
\label{slope}
\end{figure}

The frequency shift increases sharply with frequency and degree.  Although
several authors multiply the frequency shift by the mode inertia to remove its
$\ell$ dependence, we found,
as others have at lower and medium-$\ell$ \citep{Cha01}, 
that the mode inertia normalized by the inertia of
a radial mode of the same frequency gives a better agreement between the
medium and high-$\ell$ set shifts, as shown in Fig.~\ref{slope}.  The
estimated minimum-to-maximum frequency shifts can be described according to a
simple power law, \ie:
\begin{equation}
 \delta\nu^e_{n,\ell} = C_{\gamma} ~ \frac{(\nu_{n,\ell})^{\gamma}}{Q_{n,\ell}}
   \label{eq:plaw}
\end{equation}
where $Q_{n,\ell}$ is the relative mode inertia, \ie:
\begin{equation}
  Q_{n,\ell} = \frac{I_{n,\ell}}{I_{n,0}}
\end{equation}
and $I_{n,\ell}$ is the mode inertia calculated from Christensen-Dalsgaard's
model $S$ \citep[see][]{CD96}.
The frequency dependence of the frequency shift scaled by
$Q_{n,\ell}$ shows the relative importance of the contribution
from the surface layers in relation to the the solar interior
\citep{CD89}.

We fitted Eq.~\ref{eq:plaw} to estimate $\gamma$, using a weighted
least-squares minimization, as illustrated in Fig.~\ref{slope}.  For $p$
modes, we found very similar results for the medium- and high-$\ell$ sets:
$\gamma_p = 3.64 \pm0.02$ (dashed) and $\gamma_p = 3.62\pm0.04$ (solid)
respectively.
%
%
Note that in the lower panel there is a step in the frequency shift around 2.3 mHz
and only modes with $\nu \ge$ 2.5 mHz were included in the fitting of the medium-$\ell$ set
mentioned above. Despite the step, modes with $\nu <$ 2.23 mHz 
show a similar value of $\gamma$ as for $\nu \ge$ 2.5 mHz:
$\gamma_p = 3.5 \pm0.3$ (dotted and dashed line).
It is interesting to
point out that the height of the upper turning point decreases very fast with
mode frequency for $\nu <$ 2.3 mHz \citep[see Fig.2 in][]{Cha01}.
The $f$ modes for the high-$\ell$ set are strongly correlated with the solar cycle
(see lower right panel in Fig.~\ref{corr-freq})
and exhibit a behavior similar to the $p$ modes with a positive $\gamma_f$
but with a $\gamma_f = 1.6\pm0.2$
(dotted line in the upper panel of Fig.~\ref{slope}),
less than half of $\gamma_p$,
which could be an indication that different physical effects are responsible
for the solar-induced frequency shifts \citep[see][]{DG05}.
There are no $f$ modes in the medium-$\ell$ set with a correlation higher than 0.8
for comparison.

\begin{figure}[p]
\begin{center}
\includegraphics[width=10cm]{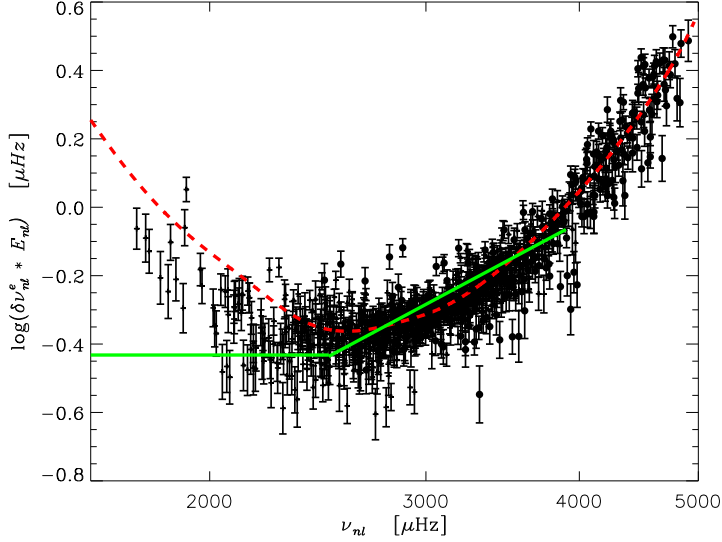}
\end{center}
   \caption{ Minimum-to-maximum frequency shift multiplied by the 
mode inertia as a function of frequency.  The medium and high-$\ell$ $p$ modes are
represented by crosses and circles respectively. The dashed line corresponds to 
the dashed line in the lower panel of Fig.~\ref{slope}. The solid line is given by 
Eq.~\ref{eq:Cha} using the values estimated by~\citet{Cha01}:
$\alpha = 0$ for $1.6 < \nu < 2.5$ mHz and $\alpha = 1.92$ for
$2.5 < \nu < 3.9$ mHz; the fitted values for $C_\alpha$ were not given
by the authors hence ad hoc values were used.
}
\label{slopeInl}
\end{figure}

\citet{Cha01} scaled the frequency shift by the mode inertia, \ie:
\begin{equation}
\delta\nu^e_{n,\ell} = C_{\alpha} ~ \frac{(\nu_{n,\ell})^{\alpha}}{I_{n,\ell}}
   \label{eq:Cha}
\end{equation}
and estimated $\alpha$ for $4 < \ell < 150$ using GONG data.  They found
$\alpha \approx 0$ for $1.6 < \nu < 2.5$ mHz and $\alpha = 1.92 \pm 0.03$ for
$2.5 < \nu < 3.9$ mHz.
To compare our results, we 
multiplied our fit, $C_{\gamma} ~ \nu^{\gamma}$, by $I_{n,0}$ and fitted 
the function $C_{\alpha} ~ \nu^{\alpha}$.
Our results using MDI medium and
high-degree modes agree well in the high-frequency range, \ie: $\alpha = 1.97
\pm 0.02$.
However, MDI medium-$\ell$ modes exhibit a negative slope $\alpha = -3.05 \pm
0.01$ for $\nu < 2.5$ mHz.  Note that 
$\log(\delta\nu^e_{n,\ell} ~ I_{n,\ell})$ 
display a quadratic dependence on 
$\log(\nu_{n,\ell}$), 
with an inflection point at 2.59 mHz 
as shown in Fig.~\ref{slopeInl}. 
This minimum in the frequency shift scaled by the mode inertia $I_{n,\ell}$ is 
also seen in Figures 4 and 5 in \citet{Howe02}.

The solar-cycle correlation coefficients decrease sharply for modes with $n >
6$ and frequencies larger than $\sim$ 4.8 mHz, some of them becoming even
negative.  However, these ``anti-correlation'' coefficients are not large
enough to draw any conclusion other than these modes present no significant
correlation with the solar cycle.

\section{Influence of Image Defocus on the Frequency Shifts}

\begin{figure}[p]
\begin{center}
\includegraphics[width=12cm]{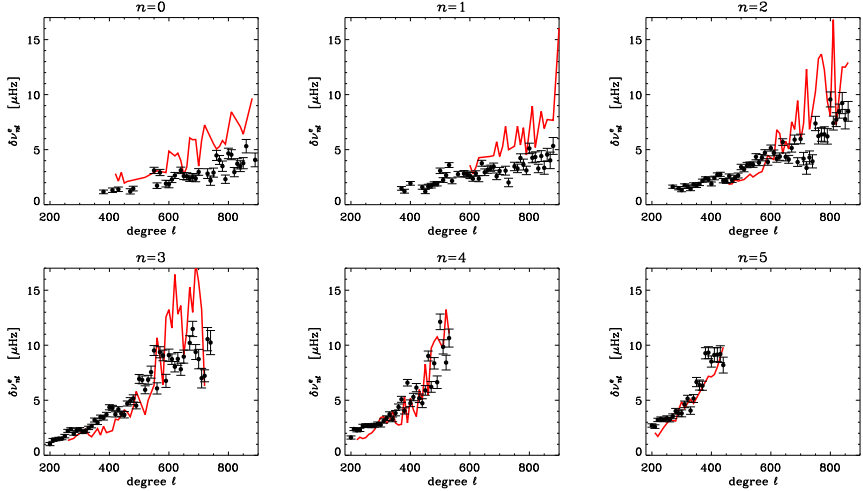}
\end{center}
   \caption{ Minimum-to-maximum frequency shift estimated using 1999 to 2004
data sets with respect to 2005 (black circles - same as in Fig.~\ref{slope})
and using 1996 to 1998 data sets with respect to 2005 (solid line).  Note
that, for the later, there is a larger scatter and the standard deviation is
also larger (not plotted for clarity) due to the small number of available
epochs.  Modes whose Pearson correlation coefficient is smaller than 0.8 or
whose slope uncertainty is larger than $20\%$ were excluded from the plot.  }
\label{comp_slope}
\end{figure}

As mentioned before, the frequency shifts for 1996 to 1998 are biased due to
the change in the amount of defocus.
To estimate the resulting bias, we computed the correlation versus the solar
activity index and estimated the minimum-to-maximum frequency shift, $\delta\nu^e$,
as described in the previous section, but using only these three epochs.
These biased frequency shifts remain strongly correlated with the solar cycle, but
the result is affected by the defocus, as illustrated in
Fig.~\ref{comp_slope}.
The differences are substantial for $n < 4$ and for $\ell > 600$.
Note that the change in the amount of defocus with respect to 2005 averages to
0.16 (in focus steps) for the later years (1999 to 2004), while that same
quantity averages to 1.77 for the early years (1996 to 1998), namely an order
of magnitude larger.  The remaining frequency bias due to change in defocus
could be as large as the frequency shift due to solar activity.

We can not discard the fact that the frequency shifts in Fig.~\ref{slope}
using 1999 to 2005 may have some significant, although very small, bias due to
variations in the defocus between the data sets (\ie\ 1999 to 2004 with
respect to 2005).
The instrumental PSF must be taken into account in the analysis at least
for data sets taken when the instrument was out of focus.

\section{Conclusions}

We analyzed the variation of the solar acoustic frequencies during solar cycle
23 using medium and high degree modes. Our high-$\ell$ mode analysis
incorporates most observational and instrumental effects that affect
specifically high-$\ell$ in the spherical harmonic decomposition to produced
unbiased mode estimates from ridge fitting.

The frequency shift scales very well with the mode inertia normalized by the
inertia of a radial mode of the same frequency and follows Eq.~\ref{eq:plaw},
where $\gamma = 3.63 \pm 0.02$ for $p$ modes using medium-$\ell$ (MDI
Structure Program) and high-$\ell$ (MDI Dynamics Program) data and $\gamma =
1.6 \pm 0.2$ for the $f$ modes, which exhibit a different behavior.

The frequency variation does not appear correlated with solar activity for
frequencies larger than $\sim4.8$ mHz. There is also an abrupt step in the
scaled frequency shift for frequencies smaller than $\sim2.3$ mHz.

We also analyzed the effect of defocusing on the high-$\ell$ frequency
determination.  This is particularly important for the first three years of
MDI observations when the instrument was set to a substantial defocus. This
has a larger impact on modes with $n < 4$ and $\ell > 600$.  Further progress
will be achieved by including a time varying PSF in the high-$\ell$ global
helioseismology analysis.

\section{Acknowledgments}

The Solar Oscillations Investigation (SOI) involving MDI is supported by NASA
grant NNG05GH14G at Stanford University.  SOHO is a mission of international
cooperation between ESA and NASA. SGK is supported by NASA grant NNG05GD58G.
NOAA Mg II Core-to-wing ratio data are provided by Dr. R. Viereck, NOAA Space
Environment Center.
The solar radio 10.7 cm daily flux (2800 MHz) have been made by the National
Research Council of Canada at the Dominion Radio Astrophysical Observatory,
British Columbia.
The International Sunspot Number was provided by SIDC, RWC Belgium, World Data
Center for the Sunspot Index, Royal Observatory of Belgium.
This study includes data from the synoptic program at the 150-Foot Solar Tower
of the Mt. Wilson Observatory.  The Mt. Wilson 150-Foot Solar Tower is
operated by UCLA, with funding from NASA, ONR and NSF, under agreement with
the Mt. Wilson Institute.

\newpage


\begin{table}[p]
  \begin{center}
    \caption{Observational periods, relative average solar active index and
average defocus.}
        \vspace{1em}
    \renewcommand{\arraystretch}{1.2}
    \begin{tabular}[!ht]{cccccc}
        \hline
     & {Starting Date} & {Duration} & {Solar Index} & {Average}  & {Starting Date}  \\
{Year} & {\em Dynamics}    & {\em Dynamics} & {rel.~to max.} &  {Defocus}    & {\em Structure}      \\
     & {[Month/Day]}     & {[days]}   & {[in \%]}          &             & {[Month/Day]}      \\
\hline
 1996 & 05/23 & 63 &  2$\,\pm\,$3           &  2.07$\,\pm\,$0.07 & 05/01 \\
 1997 & 04/13 & 93 &  5$\,\pm\,$3           &  2.53$\,\pm\,$0.07 & 04/26 \\
 1998 & 01/09 & 92 &  20$\,\pm\,$7          &  1.75$\,\pm\,$0.05 & 02/08 \\
 1999 & 03/13 & 77 &  40$\,\pm\,$13         &  0.1$\,\pm\,$0.4   & 02/03 \\
 2000 & 05/27 & 45 &  69$\,\pm\,$11         &  0.45$\,\pm\,$0.02 & 04/10 \\
 2001 & 02/28 & 90 &  61$\,\pm\,$14         &  0.57$\,\pm\,$0.04 & 01/23 \\
 2002 & 02/23 & 72 &  80$\,\pm\,$8          &  0.77$\,\pm\,$0.05 & 03/31 \\
 2003 & 10/18 & 38 &  51$\,\pm\,$17         &  0.92$\,\pm\,$0.01 & 10/28 \\
 2004 & 07/04 & 65 &  36$\,\pm\,$11         &  0.2$\,\pm\,$0.1   & 08/11 \\
 2005 & 06/25 & 67 &  30$\,\pm\,$9          &  0.34$\,\pm\,$0.02 & 05/26 \\
\hline
        \end{tabular}
      \label{table1}
   \end{center}
\end{table}

\end{document}